\documentclass[fleqn,usenatbib]{mnras} 
\usepackage{graphicx}	% Including figure files
\usepackage{amsmath}	% Advanced maths commands 
\usepackage{bm}  
\usepackage{graphics} 
\def\vlsr{v_{\rm LSR}} \def\Msun{M_\odot} 
\def\deg{^\circ}    
\def\Xco{X_{\rm CO}}  \def\Ico{I_{\rm CO}}   \def\Tb{T_{\rm B}}
 \def\co{$^{12}$CO  }   
 \def\coth{$^{13}$CO  } \def\coei{C$^{18}$O }   \def\cotw{$^{12}$CO ($J$=1-0) }   
     \def\htwo{H$_2$} 
\def\mum{$\mu$m}
\def\kms{km s$^{-1}$ } \def\Kkms{K \kms }
\def\red{}

\title[Molecular Outflow from M17]{Bipolar Molecular Outflow from M17} 
\author[Y. Sofue]{Yoshiaki Sofue
\\ 
Institute of Astronomy, The University of Tokyo, Mitaka, Tokyo 181-0015, Japan\\
E-mail: sofue@ioa.s.u-tokyo.ac.jp
} 
\date{Accepted; Received YYY; in original form} 
\pubyear{2021}
   
\begin{document} 

\maketitle
 
\maketitle  
%%%%%%%%%%%%%%%%%%%%%%%%%%%%%%%%%%%%%%%%%%%%%%%%%%%%%%%%%         
\begin{abstract}    
Kinematics of the molecular clouds in the star forming complex M17 is studied using the high-resolution CO-line mapping data at resolution ($20'' \sim 0.2$ pc) with the Nobeyama 45-m telescope. The northern molecular cloud of M17, which we call the molecular "lobe", is shown to have an elongated shell structure around a top-covered cylindrical cavity. The lobe is expanding at $\sim 5$ \kms in the minor axis direction, and at $ \sim 3/\cos \ i$ \kms in the major axis direction, where $i$ is the inclination of the major axis. The kinetic energy of the expanding motion is on the order of $\sim 3\times 10^{49}$ ergs. We show that the lobe is a backyard structure having the common origin to the denser molecular  "horn" flowing out from NGC 6618 toward the south, so that the lobe and horn compose a bipolar outflow. \red{Intensity distributions across the lobe and horn show a double-peak profile typical for a cylinder around a cavity. Position-velocity diagrams (PVD) across the lobe and horn exhibit open ring structure with the higher and/or lower-velocity side(s) being lacking or faded. This particular PVD behavior can be attributed to outflow in a conical cylinder with the flow velocity increasing toward the lobe and horn axes. }
\end{abstract}    
 

\section{Introduction} 
 
M17 is one of the most extensively studied nearby star forming (SF) regions with the wealth of observational data of molecular clouds in comparison with the distributions of young stellar objects \citep{1976ApJS...32..603L,2009ApJ...696.1278P,2018PASJ...70S..42N,2003ApJ...590..895W,2020ApJ...891...66N,2019PASJ...71S...3N,2011ApJ...733...25X,2021PASJ...73S.300K}.
Firm distance of $2.0\pm 0.2$ kpc has been obtained by trigonometric measurements of the maser sources \citep{2011ApJ...733...25X,2016MNRAS.460.1839C}. 

Nishimura et al. established a landmark view of the molecular gas in the region based on the FUGIN survey \citep{ume+2017}, and proposed triggered formation of the massive stars by a cloud collision  \citep{2018PASJ...70S..42N}. 
Besides collision scenario, there have been proposed a variety of SF mechanisms, often reaching controversial scenarios even for M17, such as the stochastic fragmentation of a cloud, sequential star formation triggered by the preceding HII region, and compression by focusing implosion of shock waves
\citep{1977ApJ...214..725E,2008ApJ...686..310H,2007ApJ...660..346P,2003ApJ...593..874T,2003ApJ...590..306D,1991ApJ...374..533L,1987A&A...171..252R,1986ApJ...307..649J,1984A&A...136...53F,2009ApJ...696.1278P,2020PASJ...72...21S}.

Regardless the SF mechanism, once the stars are formed, feedback to the molecular environment causes intriguing effects, which can be categorized as follows:
1) Further sequential triggers of SF propagating into the parent cloud \citep{1977ApJ...214..725E,2019PASJ...71S...4S}; 
2) Outflow of the gas from the SF region, which destroys the molecular clouds and excites interstellar turbulence
\citep{2019A&A...628A..21W,2018MNRAS.477.4577S,1985MNRAS.216..761C,2018MNRAS.477.4577S,2013A&A...552A..30C,2007A&A...465..931N,2009ApJ...696.1278P,2008A&A...484..361Q}. 
 
This paper focuses on the second topic of the feedback by the OB cluster NGC 6618 to the surrounding molecular clouds in the form of gas outflows.
{We extend the classical views so far drawn for the HII gas jet phenomena associated with M17
\citep{1985MNRAS.216..761C,1986ApJ...307..649J,2007ApJ...658.1119P}, but linking them to the bipolar molecular outflow toward the northern cloud.}
We assume that the star formation took place by some of the above mechanisms.

For the analysis, we use the CO-line cube data from FUGIN (Four-beam receiver system Unbiased Galactic-plane Imaging survey with Nobeyama 45-m telescope) \citep{ume+2017,mina+2016}, which had high-spatial ($20''$) and velocity ($1.2$ \kms) resolutions.

\section{Molecular Lobe and Horn}

\subsection{CO intensity maps}

\begin{figure}     
 	\begin{center}        
\includegraphics[width=7cm]{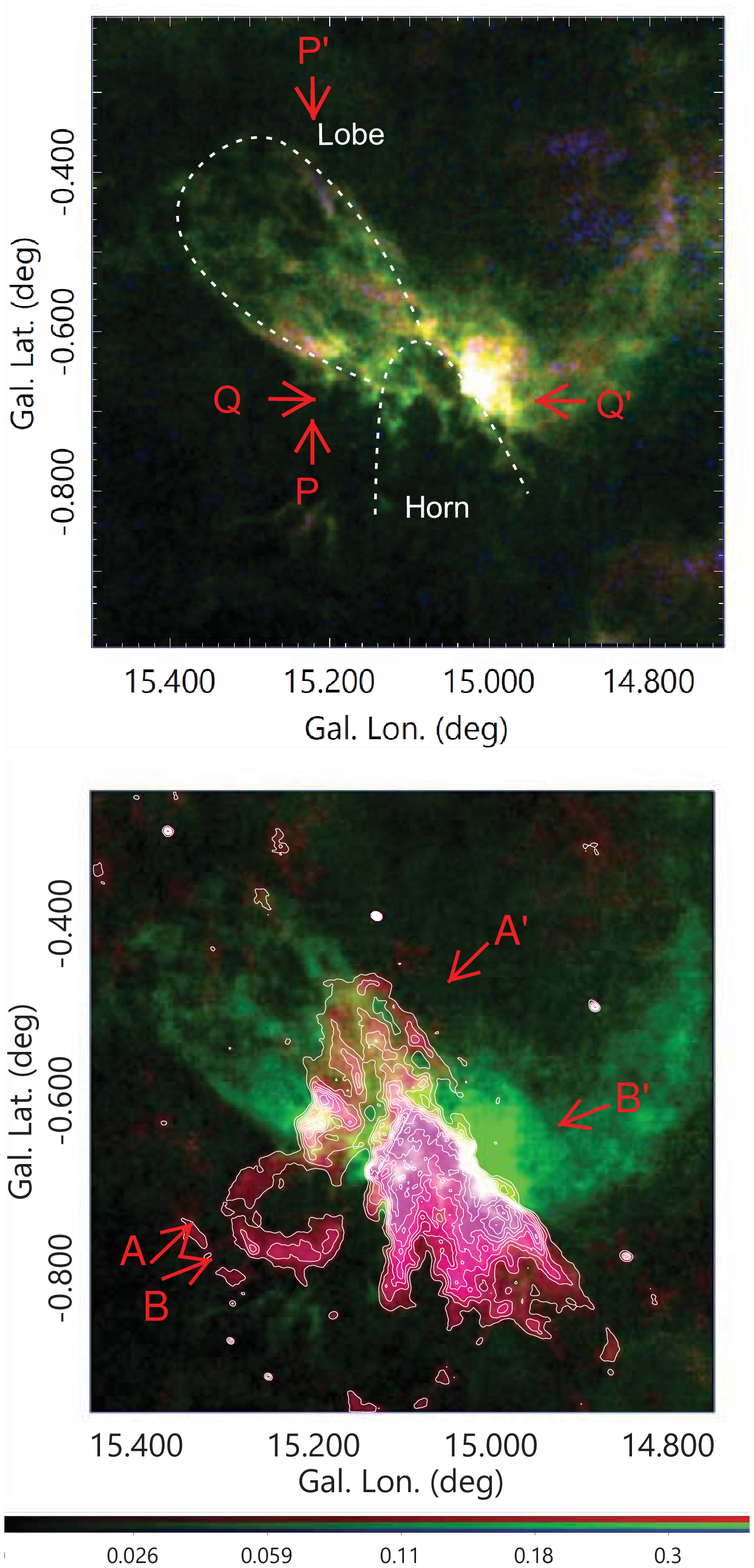}    
 \end{center} 
\caption{
[Top] $\Ico$ integrated from $\vlsr=0$ to 30 \kms in R (red) color for  \coth line, G (green) for \cotw, and B (blue) \coei. Outlines of the molecular "lobe" and conical "horn" defined in this paper and discussed are indicated by the dashed lines. 
Arrows P-P' and Q-Q' indicate approximate directions of the position-velocity diagrams in Fig. \ref{pv-lobe-horn}.
[Bottom] $\Ico$  \co line in green scale superposed by 90-cm radio continuum brightness with contours from 0.05 Jy beam$^{-1}$ with increment 0.1 Jy beam$^{-1}$. 
A thermal radio emitting shell is seen centered on  G15.25-0.75,
The arrows indicate positions of cross sections for Fig. \ref{crosssection}. 
}
\label{COradio}    
\end{figure}   

Fig. \ref{COradio} (top panel) shows a composite color-coded intensity map of the integrated intensity, $\Ico$, in the \co line by green (G), \coth by red (R), and \coei by blue (B) integrated from $\vlsr=0$ to 30 \kms.
The densest molecular cloud near G15.1-0.65 exhibits concentric arch-shaped structures around a cavity embedding the HII region M17 associated with jet-like flow toward the south \citep{1985MNRAS.216..761C}. 
Hereafter, we call the molecular arches surrounding the cavity the conical horn, or simply the "horn", after its apparent shape.

Another prominent structure is the northern molecular cloud extending in the direction at position angle $A=50\deg$ with major and minor axial lengths of about $a=0\deg.4$ (14 pc) and $b=0\deg.18$ (6.3 pc), respectively.
Hereafter, we call this cloud the northern molecular lobe, or simply "lobe".
The lobe is superposed by a number of bubbles (or loops, shells), which have major-axis diameters of $\sim 0\deg.05$ (2 pc) extending in the same direction as the lobe axis.

Integrating the \cotw intensity in the region north to the line across G15.1-0.6 at PA$=50\deg$, the molecular gas mass of the lobe is calculated to be 
$M_{\rm lobe}\sim 3.3 \times 10^4 \Msun$ for a conversion factor of 
$\Xco=2.0 \times 10^{20}$ \htwo cm$^{-2}$ [K \kms]$^{-1}$ for usual interstellar clouds
\citep{2020MNRAS.497.1851S} and mean molecular weight of $\mu=2.8$ including the metals.

The southern dense clouds close to M17 compose a molecular conical {horn} structure around a cavity open toward the south.
The {horn} exhibits a parabolic shape with the focus at G15.07-0.7 coinciding with the OB cluster NGC 6618, at which the cavity's diameter is $\sim 0\deg.1$ (3.5 pc). 
The molecular gas mass of the {horn} region of diameter $\sim 0\deg.2$ (7 pc) is estimated to be $M_{\rm horn}\sim 4.2 \times 10^4 \Msun$.

These masses are consistent with the current measurement from the same data of $3.5 \times 10^4 \Msun$ for the lobe, and $6.7\times 10^4 \Msun$ for the horn, using LTE (local theramal equilibrium) for \coth line \citep{2018PASJ...70S..42N}. 
The difference in the masses for the horn region is due to slight difference in the integrated areas.

\subsection{Molecular cavity guiding HII flow}
 
Fig. \ref{COradio} (bottom panel) shows brightness distribution of the 90-cm radio continuum emission \citep{2006AJ....131.2525H} superposed on the integrated intensity of \co line in green scale.
Three radio structures are prominent:

(i) Southern jet: The brightest structure is the well-known fan-shaped conical structure open toward the south from M17 at position angle PA$\sim 220\deg$ in the Galactic coordinates
\citep{1984A&A...136...53F,1985MNRAS.216..761C,1986ApJ...307..649J,2007ApJ...658.1119P}.
This is the HII region excited by the OB cluster NGC 6618 located at  G15.1-0.7, and is approximately embedded inside a molecular cavity concentric to the OB cluster NGC 6618
\citep{1984A&A...136...53F}.

(ii) Northern jet: A complimentary radio structure is found as a fan-shaped jet extending to the north, overlapping with the northern molecular lobe. 
This feature is mirror-symmetric to the radio {horn} in the south, making bipolar radio continuum jets.

(iii) Radio shell: The third object is the radio loop centered on G15.25-0.75 at 90 cm, which is also bright in infrared emission at 24 \mum \citep{2009ApJ...696.1278P}.
The loop has flat radio spectrum of $\alpha \sim +0.02$ between 20 and 90 cm indicative of the thermal origin. This feature is out of our discussion.

\subsection{Bipolar nature of the lobe and horn}

 	\begin{figure}     
 	\begin{center}      \includegraphics[width=7cm]{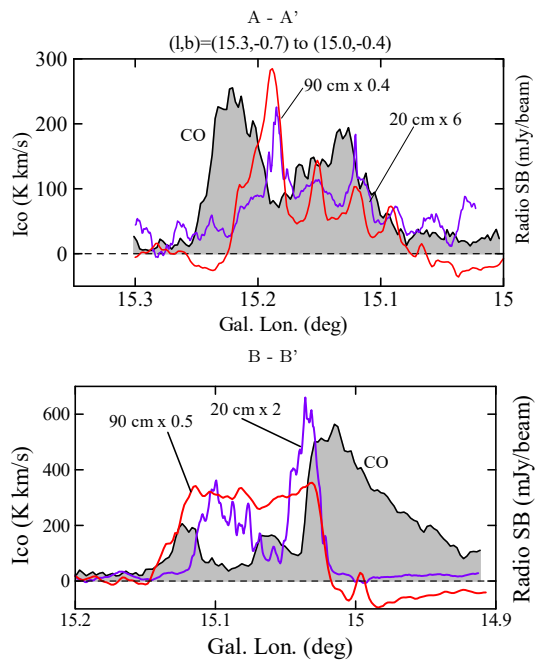}   
 \end{center}
\caption{[Top] CO (black) and 90-cm (red) and 20-cm (violet) radio continuum intensities (Helfand et al. 2006; Churchwell et al. 2009) 
along the line from $(l,b)=15\deg, -0\deg.7)$ to $(15\deg.0, --\deg.4)$ (A - A' in Fig. \ref{COradio}), showing double-peaks typical for cylindrical density distribution.
[Bottom] Cross sections of the \co intensity (shadowed line) and radio continuum across M17 {horn} along the line B - B'. }
\label{crosssection}   
\label{fig2}
\end{figure}  

The northern molecular lobe is associated with radio continuum emission, having a similar {horn} shape to that in the south.
This suggests that the lobe and horn compose a bipolar outflow.  
In order to examine the spatial relationship of the molecular and radio structures, we show in Fig. \ref{crosssection} cross sections of the \co intensity and radio brightness distributions \citep{2006AJ....131.2525H} across the lobe and horn along the lines indicated by arrows in Fig. \ref{COradio}.

The $\Ico$ cross section of the molecular lobe (top panel) exhibits a double-peak feature, typical for a cylindrical distribution of emitting gas.
The cross sections at 20 and 90 cm show similar behaviors, but are located inside the molecular double peaks.
This may indicate that the ionized gas responsible for the radio jet is embedded inside the cavity of the molecular lobe.
Almost the same structures are seen in the radio and CO cross sections across the molecular {horn} to the south near M17 (bottom panel).

Thus the cross sections in radio continuum and CO intensities are similar, and the similarity is seen both in the northern lobe and southern horn.
These similarities suggest a common origin of the lobe and {horn} related to the energy injection by the OB cluster NGC 6618 located at the root of the lobe and horn.

\subsection{Lobe and bubbles in unsharp-masked map}

 	\begin{figure}     
	\begin{center}      \includegraphics[width=6cm]{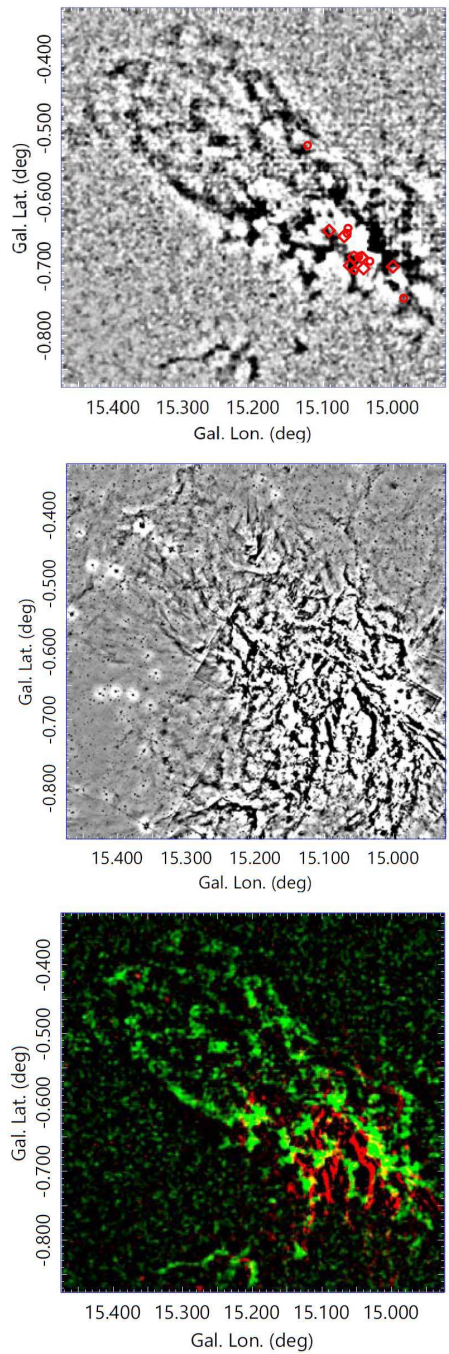}   
 \end{center}  
\caption{[Top] Unsharp-masked $\Ico$ map around M17, showing numerous bubbles. 
Red diamonds and circles indicate O stars of NGC 6618 brighter and fainter than O8, respectively (Povich et al. 2009).
Note the alignment of the O stars with the lobe's major axis.
[Middle] Same but 8 $\mu$m brightness distribution from GLIMPSE
(Chrchwell et al 2009).
[Bottom] Superposition of CO (green) and 8 \mum~ (red) unsharped images. Note the coherent alignment of the hot dust and molecular gas.
}
\label{us-bubbles}    
\end{figure}   

Fig. \ref{us-bubbles} shows unsharp-masked $\Ico$ image of the lobe and {horn}, where the oval structure of the lobe is clearly seen, suggesting that the whole structure composes an elongated large bubble.
Positions of the O stars of NGC 6618 are shown, where big diamonds are those brighter than O8 and are circles fainter O stars \citep{2009ApJ...696.1278P}.
Note that the O stars are aligned in the direction parallel to the major axis of the molecular lobe.

The lobe is further superposed by numerous small bubbles with major diameters of $0\deg.05 \sim 2$ pc, indicating that the lobe surface is patchy and bubbly.
The bubbles seem to cover the entire region including M17 and its neighboring area.

The middle panel shows an unsharp-masked image of the same region in the 8 \mum~ infrared emission from GLIMPSE \citep{2009PASP..121..213C}, showing also numerous bubbles and filaments.

While the horn region is much brighter and structural at 8 \mum~ for the higher temperature dust, bubbles and filaments of the lobe and horn appear to emanate from a common origin around NGC 6618, composing a bipolar structure.
Although bipolar, the lobe's top is closed with less number of dust filaments, while the {horn} is more open to the south with more developed filaments.
Therefore, the lobe may be an outflow with under-developing cavity, while the {horn} is a well evolved flow open to the inter-cloud space.
The reason of this asymmetry can be attributed to the eccentric location of the energy source, NGC 6618, in the molecular clouds.

The bottom panel of Fig. \ref{us-bubbles} shows superposition of the CO and 8 \mum\ images. 
In general, particularly near the horn, the molecular and warm-dust filaments
are aligned parallel to each other, while coherently displaced from each other.
The alignment suggests that the hot gas is guided by the molecular structure, and the molecular gas is also affected by the high-speed flow of hot gas.

\section{CO-line Kinematics}

\subsection{Kinematics indicators: Moment maps and position-velocity diagrams}

 	\begin{figure}     
 	\begin{center}   \includegraphics[width=6cm]{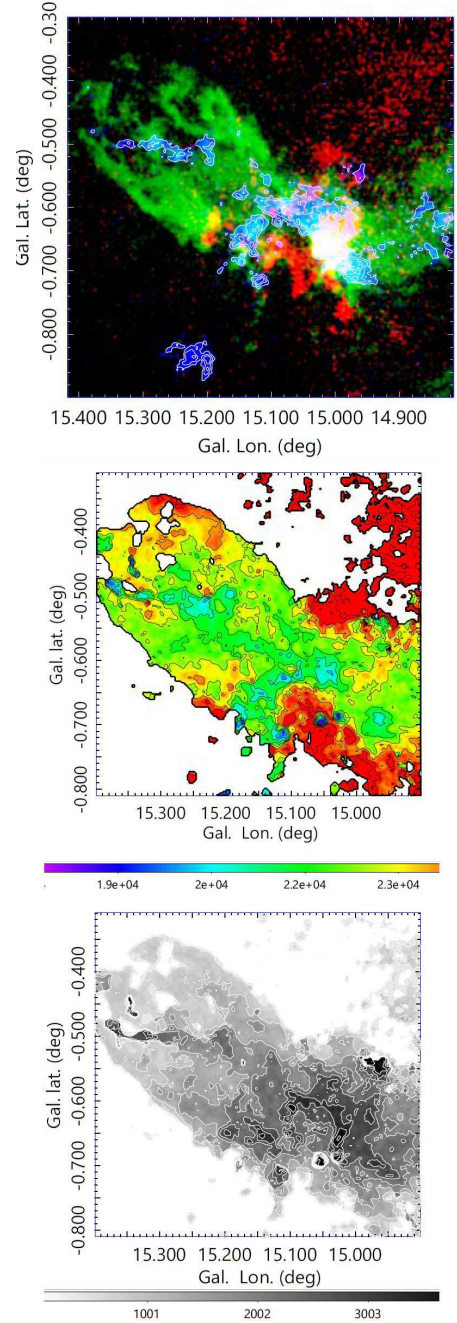}   
\end{center}
\caption{[Top] Composite channel maps at 15 \kms (B, from 5 to 15 K with contours every 10 K), 20 \kms (G, from 5 to 50 K), and 25 \kms (R, from 5 to 15 K). Note the concentric arcs
of B,G,R components with their diameters decreasing with velocity  indicative of an open conical outflow.  
 [Middle] velocity field (moment 1 in m s$^{-1}$). [Bottom] velocity dispersion (moment 2 in m s$^{-1}$).}
\label{velocompo}    
\end{figure}

The morphological relation of the CO lobe and cavity with the HII region of M17 suggests collimated bipolar outflows driven by the OB cluster NGC 6618.
More direct information about the flows can be obtained by analyzing the kinematics of the CO line emission.
Fig. \ref{velocompo} shows superposed color-coded brightness maps showing \co brightness, $\Tb$, at $\vlsr=25,~ 20$ and 15 \kms in red (R), green (G) and blue (B) intensities, respectively. 
Nishimura et al (2018) have shown that the molecular clouds are composed of three velocity components, which are the dense half arc around the main cavity at $\sim 20$ \kms (green), blue-shifted large half arc at 15 \kms, and a red-shifted small-radius arc at 25 \kms. 
In this paper, we try to explain these three components as a continuous {horn} structure.
   
The densest molecular region (green) composes a cavity concave to the OB cluster NGC 6618, and has the center velocity of $\vlsr=20.0$ \kms, showing a half arch structure open toward the south.
The red- and blue-shifted components also show concentric arcs, while the blue arc has a larger diameter and the red arc smaller.

The middle and bottom panels of Fig. \ref{velocompo} show moment 1 (velocity field) and moment 2 (velocity width) maps of the \co line emission between $\vlsr=0$ and 30 \kms, respectively.
Moment 1 map shows a velocity gradient in the direction of the major axis of the lobe in the sense that the northern top is receding (redder) from us.
Moment 2 map shows that the molecular arcs around the cavity at M17 have large velocity dispersion.  
  
Position-velocity diagram is the most useful indicator of the radial motions, cutting across the object in a plane parallel to the line of sight.

 	\begin{figure}     
 	\begin{center}  
P - P'  \hskip 4cm Q - Q'\\
\includegraphics[width=9cm]{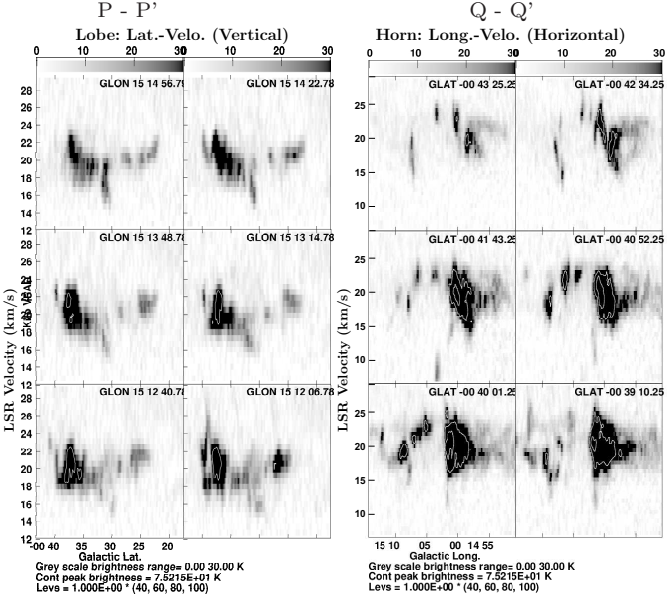} 
\end{center}
\caption{
[Left] Latitude-velocity $(b-\vlsr)$ diagrams across the lobe (vertical slices near the line P - P'ofFig. 1 oblique to the lobe axis).
\red{Note the partial ring structure.} 
[Right] Longitude-velocity diagrams $(l-\vlsr)$ (LVD; horizontal slices near line Q - Q' of Fig. 1 across the horn, showing \red{rings open in the velocity direction
}. }
\label{pv-lobe-horn}   
\end{figure}

In Fig. \ref{pv-lobe-horn} (left panel) we show typical latitude-velocity $(b-\vlsr)$ diagrams across the molecular lobe, showing vertical slices near the line P - P' in Fig. \ref{COradio}).
The right panel shows longitude-velocity diagrams $(l-\vlsr)$ (LVD) across the southern horn, showing horizontal slices near the line Q - Q' in Fig. \ref{COradio}).
The diagrams show half-ring or broken (open) ring  ( donuts) shapes in the PV plane, and will be discussed later in detail.

\subsection{\red{PV ring open in the velocity direction across the lobe}}

 	\begin{figure}     
 	\begin{center}  \includegraphics[width=7cm]{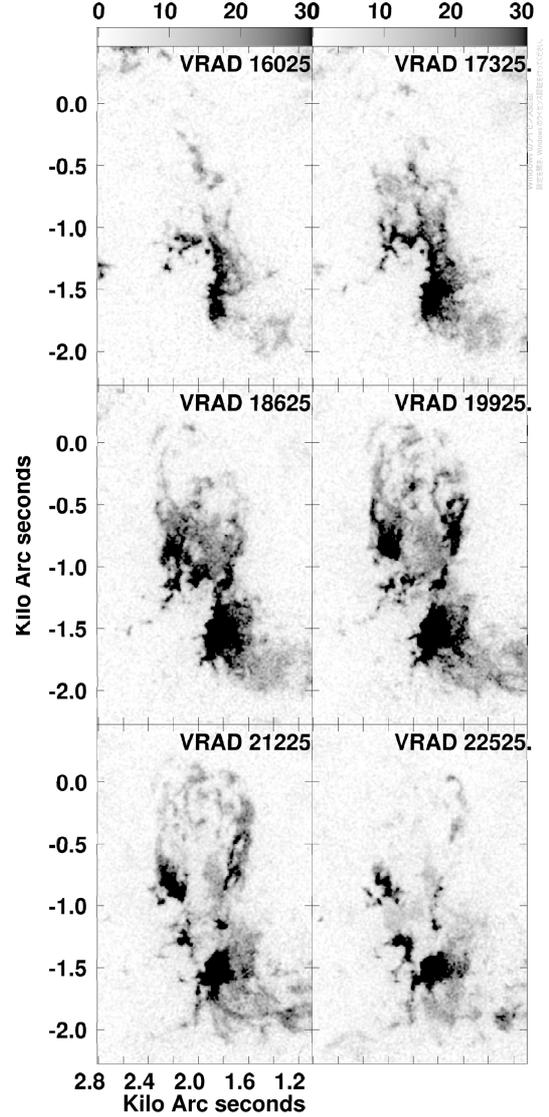}   
\end{center}
\caption{Channel maps of M17 in \co line rotated by $-50$ degrees. 
The molecular lobe extends vertically in this figure, showing a cylindrical cavity connected to the concentric cavity around M17.
Because of the coordinate rotation, the panels indicate only the relative positions by pixels of each size $\Delta X=\Delta Z=8''.5$. 
For the absolute coordinates, refer to the data presented in the other figures.}
\label{ch-rot}    
\end{figure}   

Fig. \ref{ch-rot} shows channel maps of \co brightness temperature cut out from a cube rotated by $-50\deg$ on the sky, so that the lobe axis becomes vertical in the figure.
At velocity of $\vlsr=20 \pm 0.1$ \kms, the cylindrical cavity along the major axis is well collimated between the two vertical side edges.
This cavity continuously extends to the south, and is connected to the cavity of the {horn} around M17. 
The bottom of the side edges of the lobe are also connected to the arcs of the horn.

The latitude-velocity diagrams of the lobe in Fig. \ref{pv-lobe-horn} exhibit \red{partial ring structure} open in the velocity direction.
Such curved features can be understood as due to expansion of the lobe in the near side of the cylindrical cavity at $v_{\rm exp}\sim 5$ \kms.

 	\begin{figure}     
 	\begin{center}     \includegraphics[width=8cm]{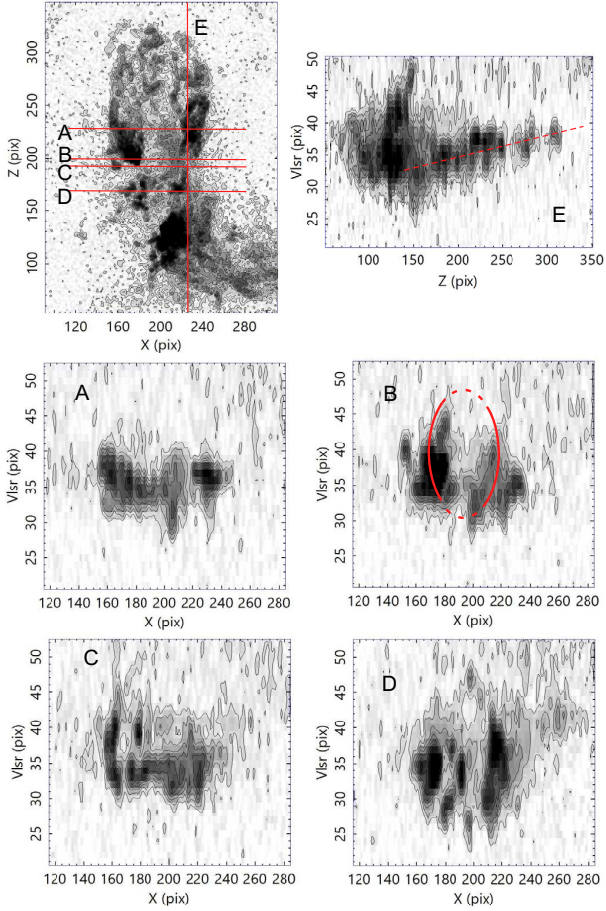}   
\end{center}
\caption{[A-D] PVDs perpendicular to the molecular lobe along the red lines in the $\Tb$ map at $\vlsr=20$ \kms at top left, \red{showing PV ring, lacking the cap and/or bottom in the velocity direction,} as indicated in panel B. 
The coordinates are rotated by $-50\deg$ with respect to the Galactic coordinates, and pixel sizes are $\Delta X=\Delta Z=8''.5$, and velocity pixel is $\Delta V=0.65$ \kms. The coordinates values are relative pixel numbers.
[E] PVD along the western edge of the lobe parallel to the axis, showing velocity gradient indicative of accelerating recession of the top.}
\label{pv-perplobe}    
\end{figure}  

\red{In Fig. \ref{pv-perplobe} we show PV diagrams  perpendicular and parallel to the lobe axis along the inserted lines in the $\Tb$ map at $\vlsr=20$ \kms in the top-left panel.}

\red{The PVD along line A shows an incomplete ring similar to those found in the $(b-\vlsr)$ diagrams in Fig. \ref{pv-lobe-horn}, indicating expanding motion of the near side of the lobe. 
PVDs along lines B, C and D show also ring, but they are open, or partially fading in the higher- and/or lower-velocity sides.} 

\red{As shown in Fig. \ref{fig2}, the intensity cross section of the lobe indicates cylindrical distribution of molecular gas.
If the cylinder is at rest or flowing at constant velocity, the PV diagram must show also double peaks at the same velocity.
If the cylinder is simply expanding, the PVD must show a closed ring, or a donuts shape, with constant intensity along the ring.  }

\red{The here observed PVD shows ring, but its  high and/or low velocity cap(s) are lacking, drawing two vertical arcs with the concave sides facing the cavity. Or, the PVD draws a barrel-shaped structure bulging out in the middle.
The arcs are sometimes associated with more extended, high-velocity filaments in the concave edge(s).
These open-ring behavior in PVD can be uniquely explained by a conical outflow with the velocity increasing toward the axis, as will be shown in the next subsection. }
 
Combining the expansion velocity, $v_{\rm exp}\sim 5$ \kms, with the total mass of the lobe ($ \sim 3.3\times 10^4 \Msun$), we estimate the kinetic energy of the expanding motion of the lobe to be $E \sim 1/2 M v_{\rm exp}^2 \sim 3\times 10^{49}$ ergs.

\subsection{\red{LV ring open in the velocity direction across the horn}}

 	\begin{figure}       
 	\begin{center}
          \includegraphics[width=5cm]{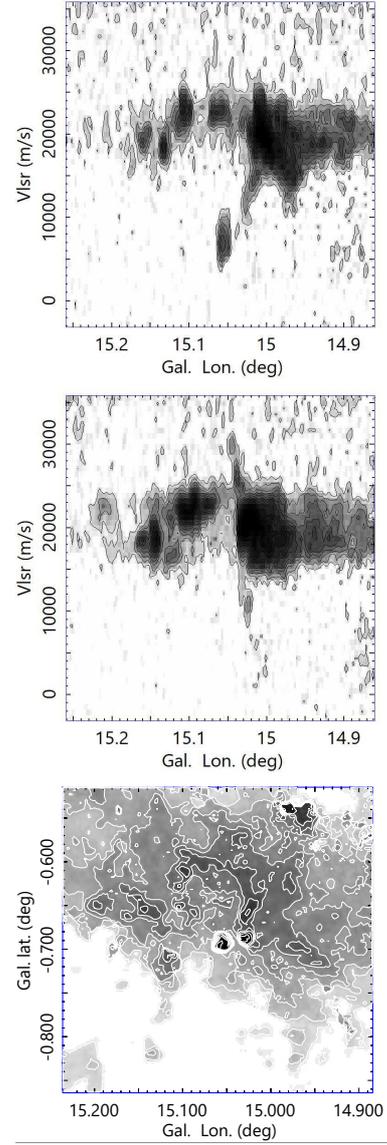}   
\end{center}
\caption{[Top] LVD along $b=-0\deg.693$ across NGC 6618, showing high-blue shifted clump at $\vlsr\sim 7$ \kms. 
Contours are from $\Tb=5$ K at 5 K interval.
[Middle] LVD at $b=-0\deg.676$, showing Open ring  and a high-velocity width skin. 
[Bottom] Velocity width (moment 2) map with contours from 0.5 \kms by 0.5 \kms interval. 
Note the high dispersion toward NGC 6618 and the {horn} wall. 
}
\label{lv-n6618}  
\end{figure}   

Kinematical property of the molecular {horn} toward the south is shown by LV diagrams in Fig. \ref{pv-lobe-horn} (right panel).
They exhibit open, conical feature symmetric to the mean velocity at $\vlsr\sim 20$ \kms at $l=15\deg.05$.
The dense molecular gas at $\vlsr=20$ \kms has a cavity of diameter  $\sim 0\deg.05$.
It is followed by an approaching flow with decreasing radial velocity up to $\vlsr = 10$ \kms, and the diameter increases with the approaching velocity.
In the red-shifted side, it shows a similar symmetric behavior in a smaller velocity range up to $\vlsr=25$ \kms.
As a whole, the LV diagrams compose \red{ring-like shape (donuts shape)} open to higher and lower velocity sides. 
\red{Fig. \ref{lv-n6618} shows closer look at LVDs near NGC 6618, showing again open rings associated with high-velocity side edges and clumps.}

%\subsection{Horn's mass and energy}
The major part of the {horn} consists of dense and massive shells of total mass $\sim 4\times 10^4 \Msun$, as calculated in the previous section.
The shells have higher velocity dispersion of $\sim 5 $ \kms than that in the surrounding molecular gas.
The kinetic energy of the {horn}  is then on the order of $E \sim 10^{49}$ ergs.

The mass of the outflow component at $\sim \pm 10$ \kms is smaller, about $\sim 1/10$ of the total horn mass.
This component has kinetic energy of $E_{\rm flow}\sim 4\times 10^{48}$ ergs.
Thus, the molecular {horn} has total kinetic energy on the order of $10^{49}$ ergs, comparable to that for northern molecular lobe.

\subsection{\red{PVD types across ring, shell, cylinder and conical horn}}

 	\begin{figure}     
 	\begin{center}      
 	\includegraphics[width=8cm]{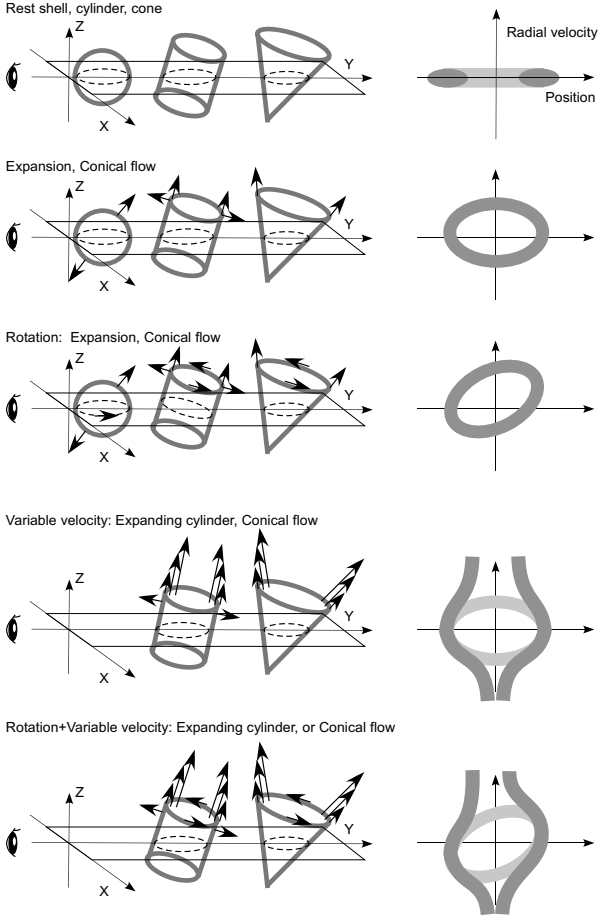}   
\end{center}
\caption{Various types of PVD patterns corresponding to shells, cylindrical flows, and/or conical flows, which have expansion, rotation, and/or variable flow velocities. }
\label{pvdpattern}   
\end{figure}   

 \begin{figure}     
 \begin{center} \includegraphics[width=8cm]{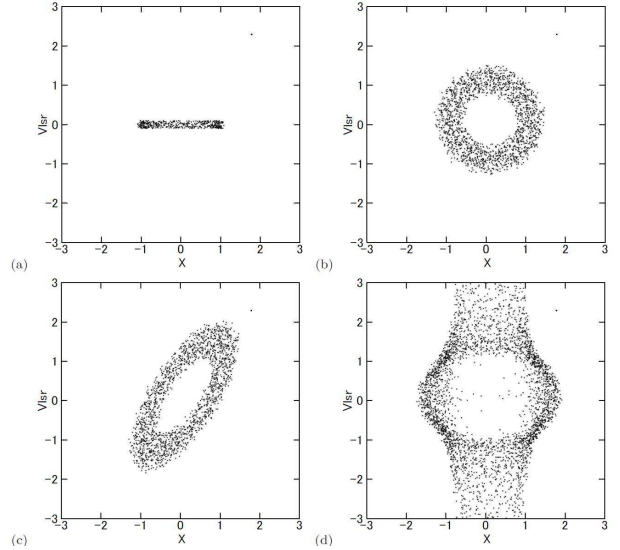}    
\end{center}
\caption{Position-velocity diagrams for:
(a) Cylinder at rest or with constant outflow velocity, or a shell at rest;
(b) Expanding ring, shell, and cylinder, as well as a conical {horn} with constant outflow velocity; 
(c) Rotating and expanding ring or cylinder;
(d) Conical cylinder with radially varying flow velocity increasing toward the axis.}
\label{pvdsimu}    
\end{figure}
Position-velocity diagrams are useful to distinguish the internal kinematical properties of interstellar clouds.
In Fig. \ref{pvdpattern} we present a gallery of various PVD behaviors corresponding to expansion, outflow, rotation, radially variable flow speed, and their combinations.
In Fig. \ref{pvdsimu} we show typical examples of model PVDs for kinematical structures as follows.\\

%\begin{itemize}
  (a) Horizontal straight ridge with the intensity attaining maximum at both edges, representing a double-peak intensity distribution.
    This feature is typical when cloud's shape is a ring, shell, cylinder, or conical horn at rest.
    
  (b) Ring or oval PVD with constant intensity. This is typical for a cloud that is a ring, shell or a cylinder, and  expanding or contracting. A conical cylinder with constant flow velocity shows also a ring/oval PVD.
    
  (c) Tilted ring PVD, showing that the cloud is a ring or cylinder, and is rotating and expanding. If there is no expansion, the diagram tends to a tilted straight line.
    
  (d) \red{Open ring} PVD, showing that the cloud is a conical cylinder with outflow velocity increasing toward the axis.
    This has been found in the M17 lobe and horn, as discussed in this paper.

\begin{figure} 
 	\begin{center}      
 	\includegraphics[width=8cm]{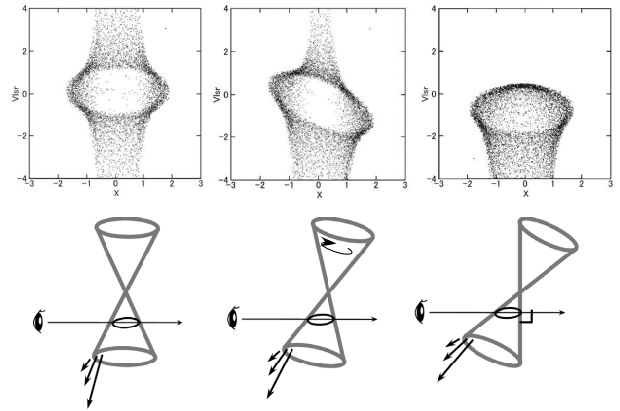}    
\end{center}
\caption{[Top] Calculated PVD crossing a conical flow with increasing velocity toward the axis, corresponding to the orientation shown in the bottom panels.
In the second panel, the {horn} is rotating, while no rotation in the left and right panels.
The simulation well reproduces the 'open PV rings' as observed in the CO lobe and {horn} in Figs. \ref{pv-lobe-horn}, \ref{pv-perplobe} and  \ref{lv-n6618}.
[Bottom] Schematics of a conical cylinder flow with increasing velocity toward the axis, and the line of sight perpendicular to the axis, oblique rotating horn, and perpendicular to a side wall. }
\label{PVbarrel}   
\end{figure} 

The here observed open \red{ring} shape of PVD cannot be explained by the currently known simple cloud's motion such as expansion, outflow, rotation, or any of their combinations.
They can be naturally understood by a model of a conical cylinder outflow with increasing velocity inside the {horn} toward the axis.
Fig. \ref{PVbarrel} shows a schematic illustration of the relation between the line of sight and horn, and resulting model PV diagrams.
Such a conical flow, whose velocity changes with radius, will be produced along the wall of a molecular {horn} due to high-speed flow of high-pressure gas generated by energy injection from massive stars formed at the root of the horn.  
Molecular gas and clumps on the horn's wall facing the cavity is accelerated by the high-velocity flow due to the friction and dynamical pressure, while the gas behind the wall (inside the cloud) remains at rest or is only weakly accelerated.

\subsection{Velocity gradient and expansion in the axis direction of the lobe}

Panel E of Fig.  \ref{pv-perplobe} shows a PVD along the western edge of the molecular lobe, parallel to the major xis, exhibiting velocity gradient of $dv/dz \sim 8.2$ \kms per degree = 0.23 \kms pc$^{-1}$ with the northern top receding from the Sun.
This can be attributed to an expanding motion of the lobe toward the north at $v_{\rm lobe top}\sim 3.2 / \cos~ i$ \kms with respect to the origin near NGC 6618 for a length of the lobe of 14 pc, where $i$ is the inclination of the axis.

\subsection{High blue-shifted velocity clump}

The top panel of Fig. \ref{lv-n6618} shows a LVD across NGC 6618 along $b=-0\deg.693$, showing an open ring or ring feature with radius $0\deg.13=4.5$ pc and half velocity width of 6 \kms.
A high-velocity blue-shifted clump is found at $\vlsr\sim 7$ \kms and $l=15\deg.06$.
This clump seems to be linked to the high-velocity width arc extending from the right-side wall of the PV ring. 
This component has been identified as the very low velocity (VLV) component by Nishimura et al (2018).

 	\begin{figure}       
 	\begin{center} \includegraphics[width=7cm]{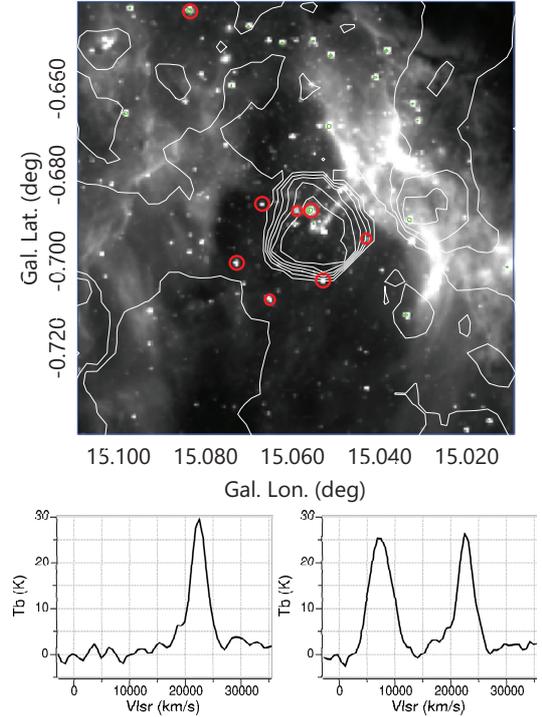}   
\end{center}
\caption{[Top] Velocity-dispersion (moment 2) map by contours  at 0.5 \kms interval as superposed on the 4.6 \mum~ image from GLIMPSE. The blue-shifted clump at $\vlsr=7$-\kms positionally coincides with the tip of an elephant trunk associated with the O stars.
Big red circles are O stars brighter than O8, and small red circles are fainter O stars of NGC 6618 (Povich et al. 2009).
[Bottom] \co line profiles toward an off-clump region (left; $0\deg.02$ to east), and toward the center of the clump (right).
}
\label{n6618+m2+4.6mu}  
\end{figure}

This 7-\kms clump, overlapped by the continuous red-shifted component in the background at $\vlsr\sim 22$ \kms, produces high-velocity width in the moment 2 map (bottom panel of Fig. \ref{lv-n6618}), resulting in the apparent large velocity width, $\delta v \sim 15$ \kms.
However, each has a normal velocity with: $v_{\rm wid}=5$ \kms (FWHM) for the 7-\kms clump, and 3 \kms for 22 \kms component (bottom panel of Fig. \ref{n6618+m2+4.6mu}).

Fig. \ref{n6618+m2+4.6mu} shows a moment 2 map around NGC 6618 superposed on a 4.6 \mum~ image from GLIMPSE \citep{2009PASP..121..213C}. 
The 7-\kms clump coincides in position with the tip of the elephant trunk associated with the central O stars of NGC 6618.
The elephant trunk is extending from the dense convex edge of the SW cloud making the horn wall facing the HII region.
The shape of the 7-\kms clump is round and intensity distribution is plateaued with the diameter of about $d=0\deg.015=0.5$ pc.

The integrated CO intensity of the clump is measured to be $\Ico\sim 65$ K \kms.
This yields molecular mass of about $M_{\rm clump}\sim 2.3\times 10^2 \Msun$ including metal.
The clump's kinetic energy due to the excess velocity of -10 \kms with respect to the LV \red{ring} center at 17 \kms is estimated to be $E_{\rm k}\sim 2\times 10^{47}$ erg.
The Virial mass is estiamted to be on the order of $10^3 \Msun$ for $r=0.25$ pc and $v_{\rm half}=v_{\rm wid}/2=2.5$ \kms.
Since $M_{\rm clump} < M_{\rm vir}$, the 7-\kms clump is not gravitationally stable.

As to the origin of the 7-\kms clump, we may consider the following mechanisms:\\
(i) The clump was accelerated by the high-speed outflow inside the horn, and is by chance oriented on the same line of sight to the OB stars.\\
(ii) The clump is located close to or in touch with the OB stars, and was accelerated by the strong UV photon pressure by the OB stars.\\
In either case, we may be observing the moment of cloud disruption by the energy injection from the OB cluster.
The disruption time is on the order of $r/v\sim 10^4$ y.  

An alternative possibility is that the clump is gravitationally bound by an interior massive object with $M\sim 10^3 \Msun$.
In this case, however, the particular location and the high blue shift remain unexplained.

\subsection{High-velocity-width wall}

{The second panel of Fig. \ref{lv-n6618} shows a LVD across the high-velocity dispersion arc along $b=-0\deg.676$, which exhibits a open ring structure open to the both velocity sides.
It has a radius of $\sim 0\deg.1=3.5$ pc and mean LSR velocity 19 \kms, respectively.
The straight and narrow vertical ridge extends for a velocity width as large as $\delta v\sim 25$ \kms, which we call the high-velocity width (HVW) component.
The average brightness of this ridge is $\Tb\sim 10$ K and the integrated intensity is $\Ico \sim 250$ \Kkms.
The longitudinal extent is as narrow as $\delta l\sim 0\deg.007 \sim 0.25$ pc, and is extended in the latitude direction for $\delta b \sim 0\deg.1\sim 3$ pc.}

The molecular gas mass of this feature is then estimated to be $M_{\rm HVWC}\sim 8 \times 10^2 \Msun$.
This molecular mass is an order of magnitude smaller than the Virial mass, $M_{\rm vir}\sim  10^{5} \Msun$, so that the clump is not gravitationally bound.
Here the Virial mass  was estimated by
$M_{\rm vir}\sim 3 r v_{\rm half}^2/G$ \citep{1987ApJ...319..730S}, where $r=(1/2)\times 3$ pc is the half size and $v_{\rm half}=v_{\rm wid}/2=\delta v/2 \sim 15$ \kms is the half velocity width.
The high-velocity width may be explained by inside molecular skin of the {horn's wall}, which was stretched and accelerated by the high-velocity flow of HII gas inside the horn.

%%%%%%%%%%%%%%%%%%%%%%%%%%%%%%%%%%%%%%%%%%%%%%%
\section{Discussion: Bipolar lobe and horn by feedback of star formation}

\begin{figure}     
\begin{center}       
\includegraphics[width=6.5cm]{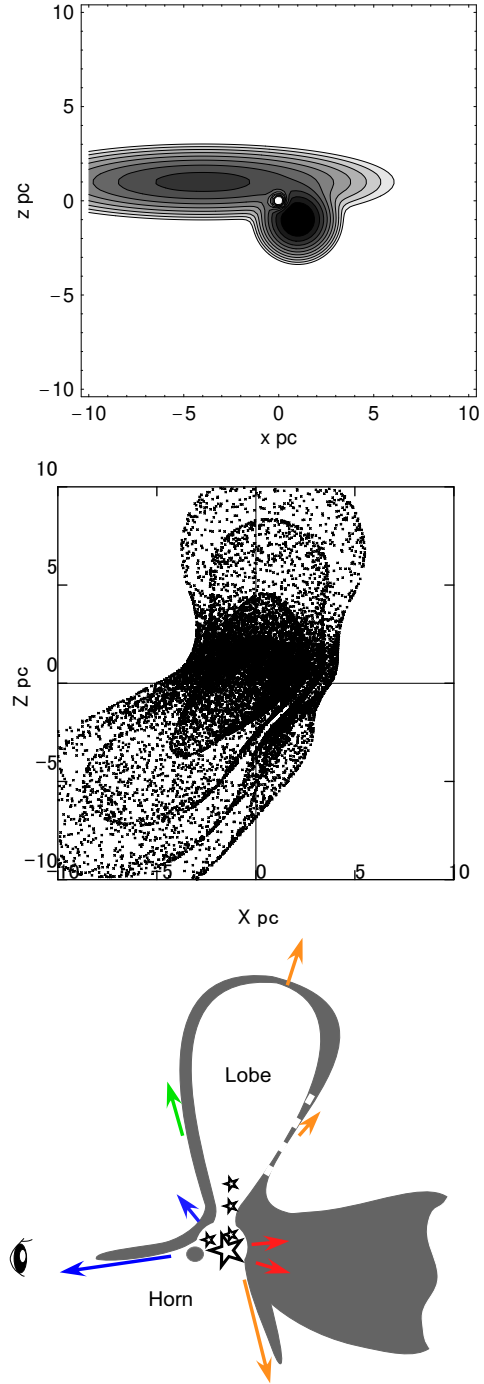}   
 \end{center} 
\caption{[Top] Two clouds between which the explosive energy was injected (white dot). 
The cloud has half sizes of 5, 5 and 1 pc in the $x,~y$ and $z$ directions, peak density of $\rho_{\rm cloud}\sim 10^4$ H cm$^{-3}$, and mass of $2.6\times 10^4 \Msun$. 
It is neighboured in touch by a higher density cloud of half sizes 2, 2 and 1 pc with peak density $5\times 10^4$ H cm$^{-3}$.
[Middle] Evolution of the shock front, forming asymmetric bipolar lobe and horn for injection energy of $E_0 \sim 10^{49}$ erg.
The shock front is drawn every $0.08$ My, and the front at $t=0.32$ My mimics the observed lobe and the horn.
[Bottom] Schematic view of the molecular lobe and {horn} around M17.}
\label{shockmodel}    
\end{figure}  

We have shown that the kinematical features observed in the PV diagrams of the \co line represent a continuous structure from the intrinsic (undisturbed) molecular cloud at LSR velocity of $\vlsr \sim 20$ \kms, on which the feedback effect of star formation (expansion and outflow) is imprinted.  
The expanding motion and conical outflow can be modeled by an explosive energy release in a molecular cloud due to newly born massive stars. 
Figure \ref{shockmodel} shows a result of calculation by applying the Sakashita's method to trace radial shock waves \citep{sakashita1971,2019MNRAS.484.2954S}.

Energy of $E_0 \sim 10^{49}$ erg was injected between two clouds in touch as illustrated in the figure (white dot), where the elongated cloud has a half-width of 5, 5 and 1 pc in the $x,~y$ and $z$ directions, and the peak density is $\rho_{\rm c1}= 10^4$ H cm$^{-3}$, and the mass $2.6\times 10^4 \Msun$. 
It is neighboured in touch by a higher density cloud with 2, 2 and 1 pc, peak density $\rho_{\rm c1}=5\times 10^4$ H cm$^{-3}$, and the mass $2\times 10^4 \Msun$.
The shock propagation is shown in the second panel, where the front is drawn every $\delta t\sim 0.08$ My.
The outermost front at $t=0.32$ My mimics the observed lobe and horn.
The calculation was performed also for continuous energy injection at a constant rate, which gave a similar result.  

The input energy, $\sim 10^{49}$ erg, was estimated from the present kinetic energy of the lobe and horn, which is as small as $\sim 10^{-3}$ times the total energy supposed to have been released by NGC 6618 in the past $\sim 0.3$ My.
We considered the seven O stars near the center of NGC 6618, of which three are brighter than O5, including O3 \citep{2009ApJ...696.1278P}.
Referring to bolometric luminosity of $10^{5.2-5.5}L_\odot$ of one O5-O7 star \citep{2010AN....331..349H}, we assume the total luminosity of the core of NGC 6618 to be on the order of $10^6 L_{\odot}$.
We, thus, estimate the total energy released in the past $\sim 0.3$ My to be on the order of $10^{52}$ erg. 
 
The shock front expands faster in the minor-axis direction of the cloud due to the rapider decrease of density.
The front then expands into an $\Omega$-shaped oval in both sides of the cloud, and further opens into the less dense interstellar space.
It finally attains a conical horn-shaped cylinder toward the bottom, and a slower expanding $\Omega$ lobe toward the top.
The top-bottom asymmetry is attributed to the eccentric position of the energy injection in the cloud.
Expansion toward right is blocked by the second cloud, resulting in right-left asymmetry producing a bent bipolar flows. 
We comment that the initial condition of the model assumes that the energy source is located at the interface of two clouds. 
It could be a subject for the future to simulate such a situation that colliding clouds form stars and accelerate a fan-jet outflow by the melon-seed effect. 

The bottom panel of Fig. \ref{shockmodel} shows a schematic view of the molecular lobe and {horn} as cut in the plane parallel to the line of sight, where radial-velocity relations are indicated by colored arrows.
The blue-shifted gas (approaching horn in the near side), green cloud (assumed to have the same velocity as the OB cluster), and red-shifted clump (receding wall of the repelled cloud in the far side) are located at different depths. 
So, red is located in the opposite side of blue with respect to green on the sky.  
The present view may be similar to that proposed for the molecular complex around the SF region W40 and Serpens cloud
\citep{2019PASJ...71S...4S}.

Although the method is too simple to discuss the details of the shock wave structure such as the density, temperature, emission, cooling, gas phases, etc.., the global behavior of the shock front can be qualitatively simulated, and is in fact mimics that obtained by the numerical hydrodynamic simulations \citep{1989A&A...216..207Y,1997A&A...326.1195C,2013MNRAS.436.3430D}.

\section{SUMMARY}
Analyzing the high-resolution CO line mapping data from FUGIN at resolution of 20 arcseconds (0.2 pc) observed with the Nobeyama 45-m telescope, we examined the kinematics of the molecular clouds in the star forming complex M17. The northern molecualr cloud, which we called the "lobe", was shown to have an elongated shell structure around a top-covered cylindrical cavity. The "horn" to the south-west exhibits open conical cylinder structure. Position-velocity diagrams of the lobe and horn showed open-ring structure, indicating expansion and outflow with the flow velocity increasing toward the axis. Kinetic energy of the lobe and horn are estimated to be $\sim 3\times 10^{49}$ and $\sim 10^{49}$ ergs, respectively. We proposed a new model that explains the morphology and kinematics of the lobe and horn by a unified view of outflow as a feedback by the OB cluster NGC 6618, and showed that they constitute a continuous bipolar structure driven by the common energy injection. 
\vskip 5mm
\noindent {\bf Acknowledgements}:
Data analysis was carried out at the Astronomy Data Center of the National Astronomical Observatory of Japan. The author thanks the FUGIN survey team for the CO-line data using the Nobeyama 45-m telescope.

\noindent {\bf Data availability}:
FUGIN CO-line data are available at URL http://nro-fugin.github.io. The MAGPIS infrared and radio continuum data were downloaded from URL https://third.ucllnl.org/gps/index.html.

\end{document}